\documentclass[preprint,aps]{revtex4}
\usepackage{epsfig}

\begin{document}

\title{Multiresolution Diffusion Entropy Analysis of time series: an application to
births to teenagers in Texas}
\author{Nicola Scafetta$^{1}$ and Bruce J. West$^{1,2}$}
\date{\today}

\address{$^{1}$ Physics Department, Duke University, Durham, NC 27708} 
\address    {$^{2}$ Mathematics Division,
Army Research Office, Research Triangle Park, NC 27709. }

\begin{abstract}
The multiresolution diffusion entropy analysis is used to evaluate the
stochastic information left in a time series after systematic removal of
certain non-stationarities. This method allows us to establish whether the
identified patterns are sufficient to capture all relevant information
contained in a time series. If they do not, the method suggests the need for
further interpretation to explain the residual memory in the signal. We
apply the multiresolution diffusion entropy analysis to the daily count of
births to teens in Texas from 1964 through 2000 because it is a typical
example of a non-stationary time series, having an anomalous trend, an
annual variation, as well as short time fluctuations. The analysis is
repeated for the three main racial/ethnic groups in Texas (White, Hispanic
and African American), as well as, to married and unmarried teens during the
years from 1994 to 2000 and we study the differences that emerge among the
groups.
\end{abstract}

\maketitle

\section{Introduction}

Time series analysis is traditionally done using linear models, such as
analysis of variance and linear regression models. Underlying these
techniques is the assumption that the phenomena of interest can be described
by a few basic standard patterns. However, the variability of the
statistical properties of a time series, such as the number of births to
married and unmarried teens shown in Figure 1, can present more complex
patterns than a periodic cycle or a linear trend, so certain precautions
must be adopted for not neglecting less evident but still important
properties of a time series. In fact, in the case of teen birth data herein
under study it was proved that after removing the background trend from the
annual periodicity, the memory remaining depends upon the marital status of
the teens \cite{scafetta,scafett2}. This fact suggests that the annual
periodicity to which human fertility is related may have social origins more
than being simply due to a natural seasonal variation and the corresponding
changes in light and temperature \cite{lammiron,rojansky}. The variability
of the data might depend on political and social changes, as well as
depending on holidays and school schedules that may induce unusual patterns
in the data.

Herein we suggest a multiresolution diffusion entropy analysis in order to
evaluate the stochastic information left in a time series after a systematic
removal, through a detrending procedure, of the memory contribution given by
identified patterns. The goal is to study a phenomenon by identifying
possible patterns, quantify the amount of information with which each of the
patterns contributes to the data and whether the identified patterns are
sufficient to capture all relevant information contained in the time series
or whether new patterns need to be found. For didactical purposes we limit
the analysis presented herein to verifying whether simple additive patterns
such as linear trends and periodic cycles are sufficient to describe the
complexity of the teen birth data. We adopt a basic regression model because
of its simple phenomenological interpretations, but we expect that the
multiresolution diffusion entropy analysis will also work with more complex
detrending models.

\section{Births to teenagers in Texas (1994-2000)}

Figure 1 shows the daily number of births to married and unmarried teen
mothers in Texas from 1994 through 2000, 2,557 days. During these 7 years,
both counts of births are characterized by annual cycles that fluctuate
around an apparent linear bias. In the case of unmarried teens this bias
increases, whereas the bias decreases in the case of married teens. The
characteristics of these data suggest that the simplest patterns can be
captured by a least-squares fitting function that involves a linear trend
plus two sinusoidal functions modulated by 1-year and 1/2-year
periodicities, respectively: 
\begin{equation}
f(t)=a+b(t-1994)+C_{1}\sin [2\pi (t-\tau _{1})]+C_{1/2}\sin [4\pi (t-\tau
_{1/2})]~.  \label{fittfunmm}
\end{equation}
The coefficient $a$ gives the approximate mean value of daily births at the
beginning of 1994, $b$ is the slope of the linear trend that gives the
annual mean increase (positive value) or decrease (negative value) of daily
births per year, $C_{1}$ and $C_{1/2}$ are the amplitudes of the 1 year and
1/2 year periodicities and, finally, $\tau _{1}$ and $\tau _{1/2}$ measure
the temporal shift of the two sinusoidal functions. The time $t$ is measured
in years. The analysis shows that during the period 1994-2000 the 99 daily
mean births to unmarried teens is almost double the daily mean births to
married teens, that being only 50 daily mean births. The number of births to
unmarried teens increases at the velocity of $b=1.92$ daily births/year,
whereas, the number of births to married teens decreases at the velocity of $%
b=-0.62$ daily births/year. Finally, the two smooth solid curves of Figure 1
clearly shows that the 1/2-year harmonic has a much more prominent influence
on the unmarried than it does on the married teens.

Table I records the fitting parameter values by repeating the same analysis
for all racial/ethnic (White, Hispanic, African American) and marital
teenagers groups in Texas during the years from 1994 to 2000. Among the
three racial/ethnic groups, the mean number of daily births to Hispanic teen
mothers is the highest with a mean of 80 births/day. The White group is
second with a mean of 44 births/day, and the African American is last with a
mean of 25 births/day. The comparison of the count of births between ethnic
groups show that the births to both White and African American teens
decreases, whereas the count of births to Hispanic teens decreases from 1994
until 1996 and increases from 1997 to 1999. Moreover,  for all racial/ethnic
groups the mean number of births to unmarried teens is almost double that
for the married teens. For Whites and Hispanics the ratio between the births
to unmarried and married young women is almost the same ($27/17=1.59$ for
White and $49/31=1.58$ for Hispanic). By contrast, for African American
teens the mean is 23 births to unmarried mothers versus only 2 births to
married mothers. The ratio is $23/2=11.5$ and is almost seven times higher
than that for the other two racial/ethnic groups. Therefore, young pregnant
African American teens are much less likely to be married than are their
White and Hispanic counterparts.

The detailed numerical analyses of the least-squares fitting curves stress
further differences among the groups. The number of births to unmarried
Hispanic teens grows with a velocity of $b=2.06$ daily births/year. In
contrast, there is only a slight increase of births to unmarried White teens
($b=0.35$), and a slight decrease in the number of births to unmarried
African American teens ($b=-0.50$). Within the married group, the number of
births to Hispanic teens is increasing slightly ($b=0.07$), the other two
groups are decreasing slightly (whites, $b=-0.65$; African Americans, $%
b=-0.04$). For all three racial/ethnic groups the unmarried teen's behavior
is always influenced by the 1/2-year periodicity. The strength of the 1-year
and 1/2-year periodicities are estimated by the intensity of the amplitudes $%
C_{1}$ and $C_{1/2}$. For the unmarried groups the two amplitudes are quite
similar, whereas, for married groups the annual amplitude $C_{1}$ is always
much stronger than the 1/2-annual amplitude $C_{1/2}$. We observe that the
annual cycle is likely related to the one year seasonal temperature/light
cycle. The 1/2-year periodicity is likely to be related to the autumn and
spring school semesters that, together with the Christmas and Summer
holidays, divide the year into two almost symmetric temporal periods. The
data present also a weekly periodicity due to the separation of the week in
working days (from Mondays to Friday) and weekends (Saturdays and Sundays)
when the hospital activity is reduced.

\section{Multiresolution diffusion entropy analysis and time series patterns}

Time series fluctuations, commonly called \textit{noise}, look random and
uncorrelated, but usually they are not. Different methods have been
suggested to extract information from the randomness of a time series; for
example, autocorrelation analysis, Hurst analysis and detrended fluctuation
analysis \cite{hurstbook,feders,dfa}, all have been used to determine the
interdependence of random fluctuations. The success of these techniques over
the last few years is due to the discovery that many natural phenomena
described by correlated fluctuations satisfy certain scaling laws \cite
{feders,west1,Mandelbrot}. A common feature of the above techniques is they
study the temporal evolution and the scaling of the variance. However, the
entropy is a more complete indicator of the stochastic information contained
in a distribution and herein it shall be prefered.

Diffusion entropy analysis (DEA) measures the time evolution of the Shannon
entropy of the probability density function (pdf) in terms of the diffusion
process generated by the time series \cite
{scafetta,nicola,scalingdetection,compressionalgorithm}. A time series of
data $\{\xi _{i}\}$ may be interpreted as the fluctuations in a diffusion
process. As in a random walk, we define trajectories by the superposition of
these fluctuations 
\begin{equation}
x^{(z)}(t)=\sum_{i=1}^{t}\xi _{i+z}~,  \label{positions}
\end{equation}
where $z=0,~1,...~$. These trajectories generate a diffusion-like process
that is described by a pdf $p(x,t)$, where $x$ denotes the variable
collecting the fluctuations and $t$ is the diffusion time. The Shannon
entropy is defined by 
\begin{equation}
S(t)=-\int_{-\infty }^{+\infty }dx~p(x,t)\ln [p(x,t)]~.
\label{diffusionentropy}
\end{equation}
DEA measures the stochastic information contained in the variability of the
smooth functions at different time scales. In fact, the diffusion
trajectories $x^{(z)}(t)$ of Eq. (\ref{positions}) generate a pdf $p(x,t)$
whose dispersion at each time $t$ is related mainly to the smooth component
of the signal characterized by the lower frequencies $f<1/t$ because the
frequencies $f>1/t$ are smoothed over by the sum (\ref{positions}).

A constant signal, equivalent to a horizontal straight line, does not have
any variability, therefore, its entropy is zero. In the presence of a
correlated noise, like the fractional Brownian motion (fGm) \cite{Mandelbrot}%
, the pdf of the diffusion process satisfies the scaling equation 
\begin{equation}
p(x,t)=\frac{1}{t^{\delta }}~F\left( \frac{x}{t^{\delta }}\right) ~.
\label{stationarycondition}
\end{equation}
Inserting Eq. (\ref{stationarycondition}) into Eq. (\ref{diffusionentropy}),
we obtain 
\begin{equation}
S(t)=\delta ~\ln (t)+A~,  \label{linearincrease}
\end{equation}
where $A$ is a constant. The coefficient $\delta $ of Eq. (\ref
{stationarycondition}) is the scaling exponent. The exponent $\delta $ can
be evaluated by using fitting curves with function of the form $%
f_{S}(t)=\delta \ln (t)+K$ that, when plotted on linear-log graph paper,
yields straight lines. For fBm the scaling exponent $\delta $ coincides with
the Hurst exponent $H$ \cite{scalingdetection}. For random noise with finite
variance, the diffusion distribution $p(x,t)$ will converge, according to
the central limit theorem \cite{Reichl}, to a Gaussian distribution with $%
\delta =H=0.5$. A correlated or \textit{persistent} noise (=tendency to
conserve the direction of the motion) would have $\delta >0.5$. An
anticorrelated or \textit{antipersistent} noise (tendency to invert the
direction of the motion) would have $\delta <0.5$.

The uppermost curves of Figure 2 show the DEA applied to the number of
births to  married and unmarried teens from 1994 to 2000. The pdf $p(x,t)$
is estimated through histograms with bin widths equal to the standard
deviation of the analyzed datasets. This has the effect of normalizing the
results. These curves do not show a real scaling behavior like Eq. (\ref
{linearincrease}) but patterns showing a short oscillation related to the
weekly periodicity and a longer oscillation of one year. The entropy grows
and suddenly decreases, reaches a minimum at one year and then increases
again. The dynamical reason for this behavior is accounted for by noticing
that a given periodicity of the data causes a periodic convergence of
distinct trajectories (\ref{positions}). After an initial spreading, with a
consequent increasing of both variance and entropy, at the end of the
oscillation there is incomplete regression back to the initial condition.

However, Figure 2 reveals much more of the time series using a
multiresolution diffusion entropy analysis \cite{nicola}. The idea is to
extract from the data the information that may be related to some noticeable
patterns in the data. In the teen birth case, a good choice is given by the
simple regression model of Eq. (\ref{fittfunmm}), that provides information
about the linear growth and the two main periodicities in the time series.
We detrend this information from the data and apply the DEA to the resulting
data set. In this way we determine the stochastic information remaining in
the detrended time series.

The ideal goal would be to find the \textit{solution} of the underlying
process, that is, an analytic function that makes the entropy of the system
vanish after the detrending procedure. A more modest goal is to extract all relevant 
information leaving only random noise. So, we first apply the DEA to the
birth data after detrending the linear ramp $y=a+bt$. The second uppermost
curves in Figure 2 show the results. We see a decrease in the entropy
because we have removed part of the information from the time series, but
the complex bending shape with the 1-year periodicity remains. The decrease
in the entropy is stronger in the unmarried than in the married teens
because their velocity of growth, $b=1.94$ births/year, is larger. The
second step is to detrend the annual periodicity (the third uppermost
curves) in addition to the linear ramp. The entropy again decreases, as
expected, but in the case of unmarried teens a clear 1/2-year cycle appears.
The figure gives a visual impression of how much more important the 1/2-year
periodicity is for the unmarried teens, than it is for the married teens. In
the third step we detrend the linear ramp plus both 1-year and 1/2-year
periodicities (the lowest curves shown in Figure 2). We continue to see the
decrease in the entropy at any point in time. This decrease in entropy is
very small for the married teens, because for them the 1/2-year cycle does
not contain very much information and, finally, we see that the weekly cycle
emerges more and more clearly.

Figure 3 shows the linear increase in entropy in logarithmic time after the
complete detrending of the deterministic process in Eq. (\ref{fittfunmm})
for all marital and racial/ethnic groups from 1994 to 2000.  It appears as
if the special condition of the diffusion entropy, described by the Eq. (\ref
{linearincrease}), has been reached. Figure 3A shows that the residual noise
is not an uncorrelated random process. In fact, the two slopes are $\delta
=0.65$ (married) and $\delta =0.60$ (unmarried), are significantly larger
than $\delta =0.5$ of the uncorrelated random noise implying that the
underlying process, after the detrending procedure, retains some anomalous
persistence. So there is still some residual memory or information that the
simple regression model described by the Eq. (\ref{linearincrease}) is not
able to extract. Moreover, we notice the similarity of the curves of Figures
3A and 3C. This may mean that the anomalous memory left in the data analyzed
in Figures 2 and 3A is due primarily to the Hispanic group. Hispanic teens
also form the largest group in terms of number of teen births for this time
period. Only in the case of married White and both married and unmarried
African American, is the slope close to the value $\delta =0.5$; a fact that
may indicate a situation of statistical randomness of the data after the
detrending process. This fact would mean that the regression model extracts
from the datasets most of the important information. Instead, for both
married and unmarried Hispanic and unmarried White teens the slope is
significantly higher than that expected for the uncorrelated random noise, $%
\delta >0.60$, indicating that the relative data show some anomalous
non-periodic persistent patterns  that the simple regression model
represented by Eq. (\ref{fittfunmm}) is not able to capture.

Figure 3B shows that the main difference between married and unmarried teens
is in the White group and indicates that births to unmarried White teens
during the period 1994-2000 were subject to some anomalous change. Finally,
Figure 3 shows differences among the racial/ethnic groups about the weekly
cycle. The likelihood of a weekday birth is greatest for Hispanics and for
Whites and slightly greater for unmarried than for married teens. African
Americans, instead, do not show the weekday preference, a fact that may
increase their odds of a neonatal death because it means that African
American young teens, for a variety of reasons, deliver their babies on
emergencies.

\section{Conclusions}

The non-stationarity of a time series is due to the presence of patterns
that need to be identified for a full understanding of a phenomenon. The
simplicity of regression models that make use of linear trends and evident
periodic cycles may capture only basic patterns of a complex system. This
fact requires the adoption of methods of analysis having the capacity of
establishing whether all relevant information is captured by a given simple
model or whether the model should be implemented with more complex patterns
that need to be identified for a better understanding of a phenomenon. The
multiresolution diffusion entropy analysis, that herein we have proposed,
has the purpose to address the above problem by evaluating the amount of
memory or stochastic information left in a time series after the systematic
removal of the memory contribution associated to the already identified
patterns.

For a didactical purpose we have applied our method to the analysis of the
number of births to teenagers in Texas from 1994-2000 for different groups.
 Table I summarizes all information that the simple regression model described by Eq. (\ref
{fittfunmm}) is able to capture. Nevertheless, figures 2 and 3 show
that such a model is not able to extract all relevant information from the
data. The diagram shown in figure 2 shows the effect on the global
information left in a time series after systematic removal of the components
of the non-stationarities associated with the regression model. This type of
diagram allows one to visually estimate the importance of every identified
pattern. Figure 3 shows that after the removal from the data of the memory
associated with the entire regression model described by Eq. (\ref{fittfunmm}%
), some datasets still retain memory patterns that manifest themselves in
the superdiffusive character ($\delta >0.5$) of the trajectories generated
by such detrended time series.

Perhaps, the implementation of Welfare Reform in Texas in the mid 1990's was
responsible for the long-range changes causing the superdiffusive behavior
with $\delta =60$ that is seen in the detrended time series of the unmarried
White teen group, but not in the married White one. This type of reform was
specifically developed to decrease non-marital child bearing and our
analysis suggests that it was particular effective for the unmarried White
teenagers. Figure 3 shows that both Hispanic and African American teenagers
do not present such a disparity between married and unmarried teens.
Therefore, perhaps the Welfare Reform in Texas was less effective for these
racial/ethnic groups. However, while the data regarding the African American
group looks random after the removal of the memory associated with the
regression model, the Hispanic group shows some anomalous residual memory ($%
\delta =0.62$ for unmarried and $\delta =0.65$ for married). Perhaps,
because this behavior regards both married and unmarried Hispanic groups,
this residual memory is a manifestation of an anomalous change of culture
and/or, as it may appear more plausible, of the Hispanic teenager population
due to immigration from 1994 to 2000 and of how this group interacts with
the social state welfare policies.

------\newline
{\ {\large \textbf{Acknowledgment:}}}\newline
The authors are grateful to Prof. Patti Hamilton of Texas Woman's University
in Denton (TX) for providing the teen birth data. N.S. gratefully
acknowledges the support from ARO grant DAAG5598D0002.



\newpage

\begin{table}[tbp]
\textbf{TABLE I}\\[1ex]
\begin{tabular}{||c||c|c||ccc||ccc||ccc||}
\hline
& All & All &  & White &  &  & Hispanic &  &  & Afr. A. &  \\ 
& Mar & UnM & All & Mar & UnM & All & Mar & UnM & All & Mar & UnM \\ 
\hline\hline
Mean & 50 & 99 & 44 & 17 & 27 & 80 & 31 & 49 & 25 & 2 & 23 \\ \hline
Max & 90 & 157 & 78 & 40 & 52 & 135 & 55 & 91 & 46 & 9 & 44 \\ \hline
Min & 21 & 55 & 16 & 3 & 6 & 39 & 11 & 21 & 9 & 0 & 9 \\ \hline
Sta Dev & 10.5 & 17.5 & 10.7 & 5.5 & 7.4 & 14.2 & 6.9 & 10.3 & 5.9 & 1.4 & 
5.6 \\ \hline
Skew & 0.10 & 0.09 & -0.02 & 0.35 & 0.16 & 0.19 & 0.15 & 0.26 & 0.23 & 0.84
& 0.24 \\ \hline
Kurt & -0.16 & -0.52 & -0.55 & 0.09 & -0.30 & -0.13 & -0.17 & -0.12 & -0.02
& 0.77 & 8e-4 \\ \hline\hline
a & 52.28 & 93.25 & 44.66 & 18.94 & 25.62 & 72.96 & 31.00 & 41.80 & 26.99 & 
1.92 & 25.01 \\ \hline
b & -0.62 & 1.94 & -0.32 & -0.65 & 0.35 & 2.11 & 0.073 & 2.06 & -0.54 & -0.04
& -0.50 \\ \hline
$C_{1}$ & 3.46 & 5.42 & 1.71 & 0.80 & 0.93 & 5.72 & 2.55 & 3.17 & 1.45 & 0.11
& 1.39 \\ \hline
$C_{1/2}$ & 0.94 & 4.48 & 1.48 & 0.32 & 1.17 & 2.80 & 0.63 & 2.14 & 1.21 & 
0.059 & 1.17 \\ \hline
$\tau_{1}$ & 0.48 & 0.52 & 0.49 & 0.46 & 0.51 & 0.50 & 0.49 & 0.50 & 0.56 & 
0.40 & 0.58 \\ \hline
$\tau_{1/2}$ & 0.49 & 0.48 & 0.50 & 0.52 & 0.50 & 0.47 & 0.46 & 0.47 & 0.47
& 0.55 & 0.47 \\ \hline
\end{tabular}
\\[1ex]
\caption{ Basic statistical analysis to the counts of births to teens in
Texas during the years from 1994 to 2000 for the three main racial/ethnic
groups -- White, Hispanic and African American -- and to the counts of
births to married and unmarried teen women. For each group we tabulate the
mean, the maximum and the minimum value, the standard deviation, the
skewness, the kurtosis and the coefficients of the least square fitting by
using Eq. (\ref{fittfunmm}). }
\end{table}

\newpage

\begin{figure}[tbp]
\epsfig{file=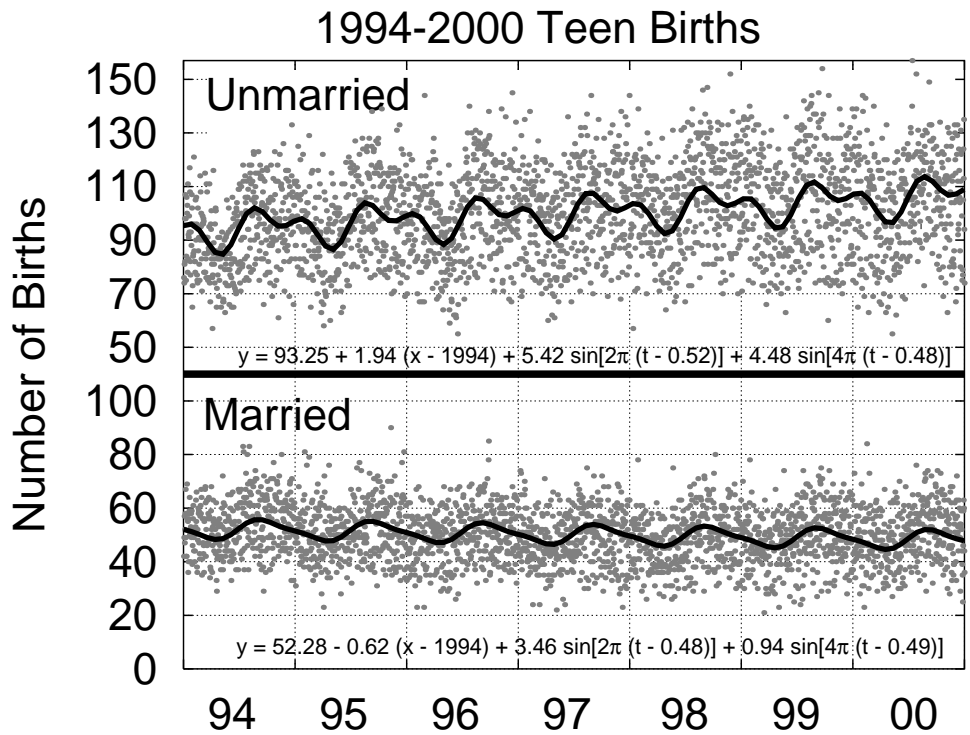, height=14cm,width=10cm,angle=-90}
\caption{Daily births to unmarried and married teens in Texas during the
years 1994 to 2000. The solid curves are obtained using the two fitting
equations at the bottom of the figure. }
\end{figure}

\newpage

\begin{figure}[tbp]
\epsfig{file=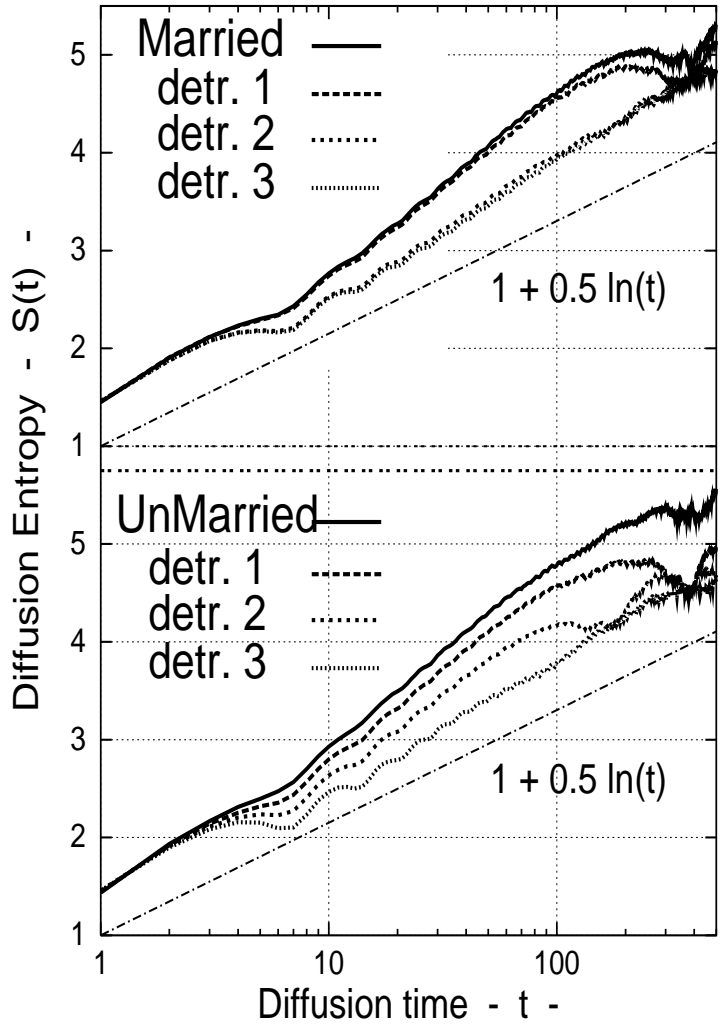,height=10cm,width=14cm,angle=-90}
\caption{Multiresolution Diffusion entropy analysis of the daily count of
births for married and unmarried teens from 1994 to 2000 at different degree
of detrending by using the linear model of Eq. (\ref{fittfunmm}). The
uppermost solid curves are the DEA of the original data without any
detrending. The dash curves show the entropy of the data detrended of the
linear ramp: $y=a+b t$. The dot curves are the DEA of the data detrended of
the linear ramp plus the annual periodicity. Finally, the lowest point
curves show the diffusion entropy of the data detrended of linear ramp plus
both 1 and 1/2 year periodicities. The straight lines indicate the slope of
the diffusion entropy of the random noise that is characterized by $%
\delta=0.5$. }
\end{figure}

\newpage

\begin{figure}[tbp]
\epsfig{file=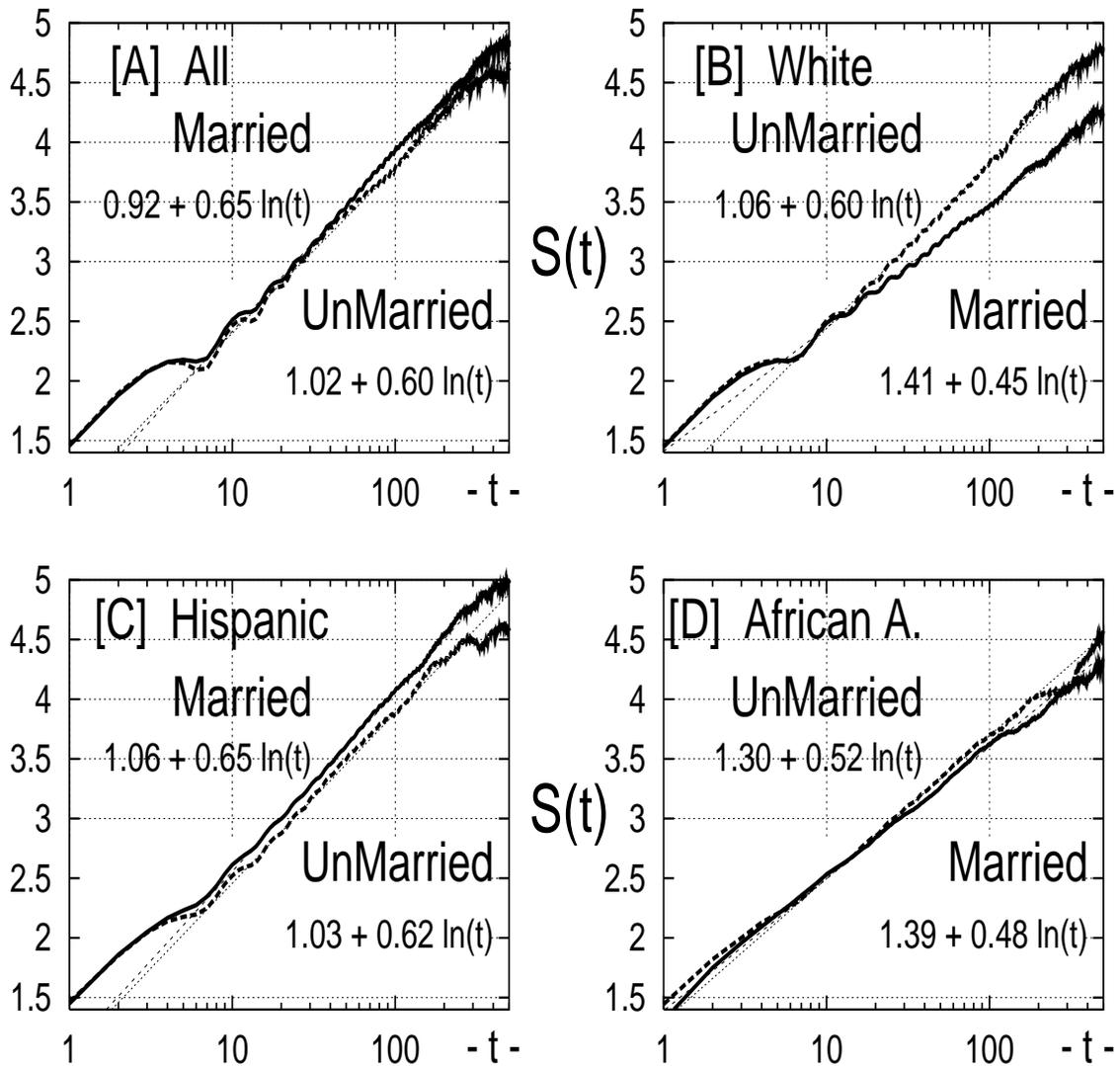,height=16cm,width=15cm,angle=-90}
\caption{Diffusion entropy analysis after the complete detrending of the
linear model represented by Eq. (\ref{fittfunmm}). Married (solid lines) and
unmarried (dash lines) teen births for the three racial/ethnic groups
(white, Hispanic and African American) from 1994 to 2000 are analyzed. }
\end{figure}

\end{document}